\documentclass{aastex}
\usepackage{emulateapj5,psfig}
\usepackage{graphicx}
\usepackage{timesfonts}
\def\arraystretch{1.25}
\makeatletter
\newenvironment{inlinetable}{%
\def\@captype{table}%
\noindent\begin{minipage}{0.999\linewidth}\begin{center}}
{\end{center}\end{minipage}\smallskip}

\newenvironment{inlinefigure}{%
\def\@captype{figure}%
\noindent\begin{minipage}{0.999\linewidth}\begin{center}}
{\end{center}\end{minipage}\smallskip}
\makeatother
\def\ltsima{$\; \buildrel < \over \sim \;$}
\def\lsim{\lower.5ex\hbox{\ltsima}}
\def\loe{\lower.5ex\hbox{\ltsima}}
\def\gtsima{$\; \buildrel > \over \sim \;$}
\def\gsim{\lower.5ex\hbox{\gtsima}}
\def\goe{\lower.5ex\hbox{\gtsima}}

\newcommand{\rc}{\rm }
\newcommand{\be} {\begin{equation}}

\newcommand{\ee} {\end{equation}}

\newcommand{\src}{1RXS\,J170849.0--400910}

\newcommand{\BSAX}{{\em Beppo}SAX} 

\newcommand{\bc}{\begin{center}}
\newcommand{\ec}{\end{center}}
\def\uu {4U\,0142$+$614}
\def\oo {1E\,1048.1$-$5937}

\def\src {1RXS\,J170849$-$400910}
\def\rx {1RXS\,J170849$-$400910}
\def\ee {1E\,2259$+$586}
 
\def \nh {N${\rm _H}$}
\def \hcm {\hbox {\ifmmode $ atom cm$^{-2}\else atom cm$^{-2}$\fi}}

\def\ltsima{$\; \buildrel < \over \sim \;$}
\def\lsim{\lower.5ex\hbox{\ltsima}}
\def\gtsima{$\; \buildrel > \over \sim \;$}
\def\gsim{\lower.5ex\hbox{\gtsima}}

\lefthead{ISRAEL ET AL.}
\righthead{THE PULSE PHASE--DEPENDENT SPECTRUM OF \src}
\submitted{Received: 16 June 2001;~~ accepted: 31 August 2001}

\begin{document}
\title{The pulse phase--dependent spectrum of the anomalous 
X--ray pulsar \src}
\authoremail{gianluca@mporzio.astro.it}

\author{G.L. Israel\altaffilmark{1,2}, T. Oosterbroek\altaffilmark{3}, 
L. Stella\altaffilmark{1,2}, S. Campana\altaffilmark{4,2}, 
S. Mereghetti\altaffilmark{5} and 
A.N. Parmar\altaffilmark{3}}

\affil{1. Osservatorio Astronomico di Roma, Via Frascati 33,
I--00040 Monteporzio Catone (Roma), Italy, gianluca@mporzio.astro.it}
\affil{2. Affiliated to the International Center for Relativistic
Astrophysics (I.C.R.A.)}
\affil{3. Space Science Department of ESA, ESTEC, P.O. Box 299, 
2200 AG Noordwijk, The Netherlands}
\affil{4. Osservatorio Astronomico di Brera, Via Bianchi 46, I--23807
Merate (Lc), Italy}
\affil{5. Istituto di Fisica Cosmica ``G.P.S. Occhialini'' del C.N.R., 
Via Bassini 15, I--20133 Milano, Italy}
\begin{abstract}
We report on the results of a 50\,ks \BSAX\ observation of 
\rx, one of the five (plus a candidate) known anomalous X--ray pulsars. 
The \BSAX\ data are consistent with a power--law plus blackbody spectral 
decomposition, making \rx\ the fourth source of this class {\rc for which 
such a spectral decomposition was found}. 
The inferred power--law slope and blackbody temperature 
are $\Gamma\sim$2.6 and kT$_{\rm BB}\sim$0.46\,keV, respectively.  
We {\rc found significant energy--dependence of the pulse profile, a 
remarkable feature for an AXP}. 
By using the power--law plus blackbody decomposition we detected 
a significant variation in at least one spectral parameter, 
the power--law photon index, as a function of the pulse phase. 
This is the first significant detection of spectral parameter variation 
in an AXP. The implications of these results are briefly 
discussed.

\end{abstract}


\section{Introduction} 
After more than 20 years since the discovery of pulsations from \ee\ 
the nature of the Anomalous X--ray Pulsars (AXPs) is still an open issue. 
Although we can be reasonably 
confident that AXPs are magnetic rotating neutron stars (NSs), their 
energy production mechanism is still uncertain. It is also unclear 
whether they are solitary objects or are in binary systems with very 
low mass companions (for a review see Israel et al. 2001 {\rc and 
references therein}).
{\rc Different production mechanisms for the observed
X--ray emission have been proposed, involving either accretion or
the dissipation of magnetic energy. Among the properties that distinguish AXPs from 
known magnetic ($\geq$ 10$^{12}$ G) accreting X--ray pulsars found in
High and Low Mass X--Ray Binaries (HMXBs and LMXBs) are: (i) spin periods in a 
narrow range ($\sim$6--12\,s), (ii) very soft and absorbed X--ray spectra, (iii) 
relatively stable spin period evolution, with long term spin--down
trend, (iv) flat distribution in the Galactic plane and three clear 
associations with supernova remnants. There are currently  
five ascertained members of the AXP class plus one likely candidate.}

The relatively bright source \rx\ was discovered early in the  ROSAT 
mission {\rc (Voges et al. 1996)}; however only in 1997 this source 
attracted much attention 
because of the ASCA discovery of $\sim$11\,s pulsations 
(Sugizaki et al. 1997). Based on the pulse period and unusually soft
X--ray spectrum the source was tentatively classified as a candidate AXP.
This interpretation was confirmed through ROSAT High 
Resolution Imager (HRI) 
observations which provided the first measurement of the period
derivative \.P$\sim$2$\times$10$^{-11}$\,s\,s$^{-1}$ (Israel et al. 1999a). 
\rx\ is one of the two AXPs for which a phase--coherent timing solution 
was obtained by a systematic monitoring program with the {\em Rossi}XTE PCA 
(Kaspi et al. 1999). The source was found to be a quite stable
rotator with phase residuals of only $\sim$1\%, i.e. comparable to or smaller
than those measured for most radio pulsars.
However, on 1999 September the {\em Rossi}XTE satellite detected a sudden
spin--up event from \rx\  which was interpreted as a ``glitch''
similar to those observed in the Vela and other young radio pulsars
(Kaspi et al. 2000). 

A search for optical counterparts in the field of \rx\ was carried   
out by Israel et al. (1999a). These authors, based on two refined ROSAT HRI 
positions, were able to rule out a massive early 
type companion (a distant and/or absorbed OB star
would appear more reddened). However, the images were taken from a
1.5\,m telescope and were not deep enough to constrain any other proposed
theoretical scenario such as a low--mass companion, a residual disk or a
magnetar. 

An association between \rx\ and the supernova remnant (SNR)
G346.6--0.2 located $\sim$12$'$ away was proposed by Marsden et al. (2001).
However, as discussed by Gaensler et al. (2001) such 
an association appears to be unlikely. An image at 1.4 GHz showed the 
presence of an arc of diffuse emission $\sim$8$'$ away from \rx\ (Gaensler 
et al. 2001), which was interpreted as a previously unknown supernova 
remnant (G346.5--0.1). Also in this case there are no convincing arguments 
for a physical association between G346.5--0.1 and \rx. No radio emission 
was detected from \rx, with upper limits of 3\,mJy on the continuum 
(5$\sigma$ at 1.4 GHz; Gaensler et al. 2001). 
In this paper we report the results obtained from a \BSAX\ 
observation of 1RXS J170849--400910 that took place before the 
``glitch''--like event detected by {\em Rossi}XTE. A two--component spectrum, 
i.e. a power--law plus a black body, was found. 
Moreover the \BSAX\ observation revealed significant energy--dependent pulse 
profile. Pulse phase spectroscopy shows that,
in the context of the blackbody plus power law spectral decomposition,
this effect is likely connected to a photon index variation. 
This observation therefore provided the first evidence for pulse 
phase--related spectral variations in an AXP. Implications of these 
results are also briefly discussed.
\begin{inlinefigure}
\psfig{figure=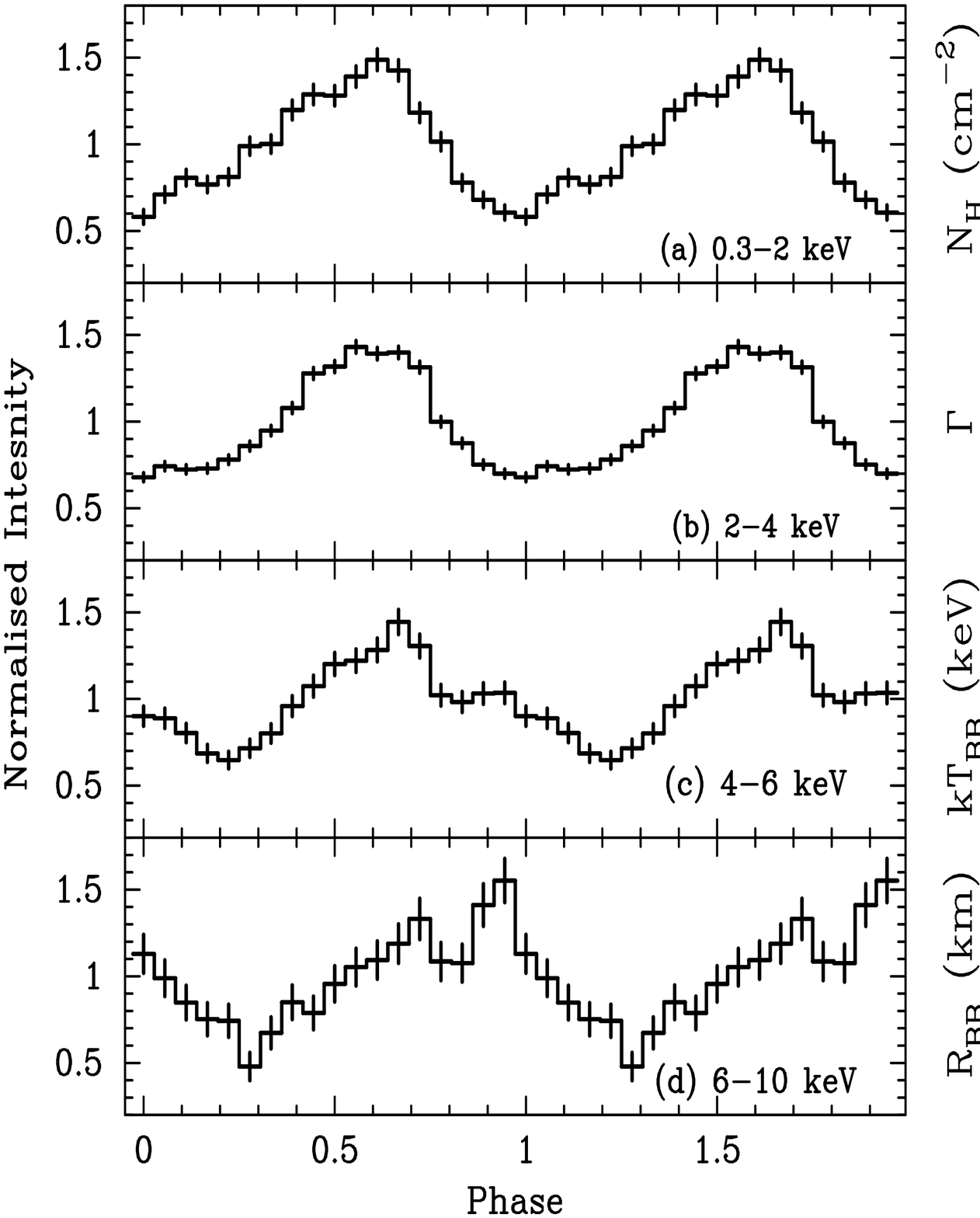,height=7.5cm} 
\caption{\rx\ MECS and LECS light curves (left panels 
labelled from (a) to (d)) folded to the best period (P=10.99915\,s) for 
four different energy intervals. For clarity two pulse cycles are shown. 
Zero phase was (arbitrarily) chosen to correspond to the minimum in the 
0.3--2\,keV folded light curve. The results of the pulse phase spectroscopy 
are shown for selected free parameters (right panels; the absorption column 
in units of 10$^{22}$). Phase intervals refer to those shown in the left panel.}
\end{inlinefigure}

\vspace{-4mm}
\section{Observations and Results} 

{\em Beppo}SAX observed \rx\ between 1999 March 31 14:27 UT and April 1 22:50 UT 
with imaging Narrow field Instruments: the Low--Energy Concentrator Spectrometer
(LECS; 0.1--10~keV; Parmar et al. 1997; 26\,ks effective exposure time)
and the Medium--Energy Concentrator Spectrometer (MECS; 1.3--10~keV;
Boella et al. 1997; 52\,ks effective exposure time). 

\vspace{-3mm}
\subsection{{\rc Timing analysis}}
The arrival times of the 0.1--10~keV photons from \rx\ 
were corrected to the barycenter of the solar system and a 1\,s 
binned light curve was accumulated. 
The MECS counts were used to accurately determine the pulse period. 
The data were divided into 12 time intervals, and for 
each interval the relative phase of the pulsations was determined. 
These phases were then fit with a linear function giving a best--fit 
period of 11.99915$\pm$0.00002\,s. 
The background subtracted light curves, folded at the best 
period in different energy ranges (Figure\,1; left panels) show an 
energy--dependent profile.  
{\rc In particular the phase interval of the minimum in the energy interval 
0.3--2\,keV folded light curve corresponds to a maximum in the 6--10\,keV 
folded light curve}.
Moreover the pulsed fraction (semi--amplitude of modulation divided 
by the mean source count rate) decreases from
$\sim$40\% to $\sim$30\% from the lowest to the highest energy band.

\vspace{-2mm}
\subsection{Spectral analysis}  
PHA spectra were obtained from the \BSAX\ position (R.A. = 
17$^{\rm h}$\,08$^{\rm m}$\,48$^{\rm s}$, Dec. = 
--40$^{\rm o}$\,08\arcmin58\arcsec; equinox 2000; 
90\% confidence level radius of 22\arcsec) of \rx\ using  an extraction 
radius of 4$'$ and 8$'$  for  the MECS and LECS, respectively.  
Background subtraction was performed using both standard blank field 
exposures and background regions taken from the observation of \rx\ 
far from the position of the AXP {\rc ending up with similar results}. 
The PHA spectra were rebinned in order to have more than 40 counts in each 
bin such that minimum $\chi^2$ fitting techniques could be reliably used. 
All those bins which were consistent with containing zero counts after 
background subtraction were 
\begin{inlinetable}
{\small
\begin{center} 
\caption{\BSAX\ phase averaged fit of \rx.} 
\begin{tabular}{lccc} 
\hline  
Spectral Parameter & PL & PL + BB & BB + BB \\ 
\hline  
 \nh (10$^{22}$ \hcm) &1.88$\pm$0.08 & 1.42$\pm$0.15&0.9$\pm$0.1 \\ 
$\Gamma$ & 3.28$\pm$0.05&2.62$\pm$0.17 &\\ 
PL flux  &4.26&3.30&\\ 
$kT_{\rm BB}$ (keV) &&0.46$\pm$0.03&0.50$\pm$0.02 \\ 
---        &&             &1.54$\pm$0.06 \\ 
BB radius (km) &&4.0$\pm$0.4&4.4$\pm$0.2 \\ 
---  && &0.30$\pm$0.02 \\ 
BB flux  &&1.2&2.8+1.3 \\ 
---                                        &&   &1.7 \\ 
$\chi^2$/dof &1.24&1.00&1.10 \\ 
$L_X$ (10$^{35}$ erg s$^{-1}$) &8.5&3.6&2.0 \\ 
  \hline  
\end{tabular}   
\end{center}} 
{\small 
\begin{minipage}{0.98\linewidth}
\textsuperscript{} ~ \\ 
NOTES --- Fluxes and luminosities refer to the 0.5--10\,keV band 
($\times$ 10$^{-11}$ erg s$^{-1}$cm$^{-2}$. 
Fluxes are not corrected for the interstellar absorption. Flux 
uncertainties are about 10\%. 
The source luminosities were derived by setting N$_H$ = 0 and assuming 
a distance of 5\,kpc.
\end{minipage}}
\end{inlinetable}

\noindent rejected. Moreover the analysis of the  MECS and 
LECS spectra was 
restricted to the 1.8--10\,keV and 0.1--5\,keV ranges, respectively, {\rc where 
the calibration of response files is more accurate}.
A constant factor free to vary within a predetermined range was 
applied in the fitting to account for known normalization differences 
between LECS and MECS.

A simple power--law model did not fit the data {\rc well} (reduced $\chi^2$ 
of 1.24 for 204 degrees of freedom, hereafter {\em dof}). 
All other single component models that we tried produced even worse results. 
A much better fit (see Figure\,2) was obtained by including a soft
thermal component, a blackbody, in addition to the power--law (reduced 
$\chi^2$ of 1.00 for 202 {\em dof}). This model was successfully fitted 
to the spectra of three other AXPs.  
An F--test shows that the inclusion of the blackbody component is highly 
significant (probability  of $\sim$6$\sigma$). The best fit was obtained 
for an absorbed ($N_H$ = (1.42$\pm$0.15)$\times$10$^{22}$\,\hcm)  
power--law with photon index $\Gamma$ = 2.6$\pm$0.2 and a blackbody component 
with temperature of $kT_{\rm BB}$ = 0.46$\pm$0.03\,keV (90\% c.l. reported; 
see Table\,1). The unabsorbed  0.5--10\,keV
flux was 1.3$\times$10$^{-10}$\,erg\,s$^{-1}$\,cm$^{-2}$. The blackbody
component accounts for about 36\% of the total {\rc absorbed} flux in the same 
band. Figure\,2 shows the spectral shape and components of \rx\ as determined 
by {\em Beppo}SAX. 
We also tried to fit the spectra using different spectral decompositions. 
Among these we find a relatively good fit (reduced $\chi^2$ 
of 1.10 for 202 {\em dof}) with a  two blackbody model (see Table\,1). 
It is worth mentioning that also in this case we obtained a soft component 
(i.e. a black body) with a characteristic temperature similar to that 
inferred with the power--law plus black body 
model. 

The data from the High Pressure Gas Scintillation Proportional Counter 
(HPGSPC) and the Phoswich Detector System (PDS) did not provide any useful 
information on 1RXS J170849--400910. In fact, due to the large FOVs of these 
instruments, the relatively short exposure time ($\sim$25\,ks), and 
steep spectrum of \rx\ the source was not detected.
\begin{inlinefigure}
\psfig{figure=f1e.eps,width=11cm,height=7.2cm} 
\caption{LECS and MECS energy spectra of \rx. The residuals 
(in units of $\chi^2$) of the best fit are also shown (see the text 
for details). The power--law and blackbody components are shown 
with dotted--dashed and dashed lines, respectively.}
\end{inlinefigure} 

\subsection{Phase resolved spectroscopy and pulse profiles}
Pulse phase spectroscopy (PPS) was carried out with the 
MECS and LECS data. 
A set of five phase--resolved spectra (phase boundaries 0.0, 0.2, 0.4, 
0.6, 0.8) were accumulated.  After rebinning and background subtraction 
these were then fit with the power--law plus blackbody model described 
in Sect.\,3.1. Due to the {\rc small} number of photons in the LECS spectra,
we removed all the PHA channels below 1\,keV. 
Initially, the blackbody temperature was fixed at its phase--averaged 
best--fit value and only the power--law parameters and blackbody normalisation 
were allowed to vary giving a cumulative (over the whole set of spectra) 
reduced $\chi^2$ of 1.10 for 466 {\em dof}. The fits were then repeated 
with the power--law component fixed and blackbody parameters free to vary 
resulting in a reduced $\chi^2$ of 1.32 for 466 {\em dof} for the best fit. 
Then all the parameters were varied and fitted together. In the latter case 
the best fit gave a reduced $\chi^2$ of 1.04 for 456 {\em dof}.   
An F--test shows that the freeing of the blackbody and 
power--law phase--averaged parameters is highly significant 
(the reduction in $\chi^2$ has a formal probability of 9$\times$10$^{-4}$ 
and 1$\times$10$^{-19}$ for the blackbody and power--law parameters, 
respectively).
In Figure\,1 (right panels) and 3 the results of the PPS are shown for 
the most interesting parameters and phase intervals. 

A check was {\rc performed} by fitting the power--law photon index values 
obtained in each phase intervals {\rc with a constant} (see right 
panels of Figure\,1). This set a $\sim$3$\sigma$ significant variation
in the  power--law photon index parameter.  
Statistical uncertainties prevent a firm detection of variations 
in the other parameters (see right panels of Figure\,1). 

The upper two panels of Fig.\,4 show the pulsed fraction versus energy 
during \BSAX\ observation for the first two harmonics. These values were 
obtained by fitting the corresponding light curves with two sinusoidal 
functions. Note that definitions of the pulsed fraction which 
involve the maximum and the minimum of the folded light curve should be 
used with care as they are dependent on the binning time and, 
therefore, the presence of unusually low or high data points. The 
sinusoidal fit, when feasible, is less sensitive to these data points. 
The first harmonic decreases from $\sim$36\% 
to $\sim$26\% as the energy increases from 0.5--2\,keV to 6--10\,keV.
A constant value of $\sim$10\% is inferred for the second harmonic. 
The last two panels refer to the ratio between the power--law and the 
total absorbed fluxes (i.e. the blackbody plus power--law spectral 
model; third panel), and the ratio between the blackbody and the 
power--law  unabsorbed fluxes (lowest panel). From 
the comparison of these quantities we can infer that: 
(i) there is evidence for a decrease of the  1$^{st}$ harmonic 
pulsed fraction at energies above 5\,keV, although the statistics are poor; 
(ii) there is evidence for an anti--correlation between the power--law 
component and the 1$^{st}$ harmonic pulsed fraction {\rc (see first and 
third panels)}, and (iii) {\rc evidence} for a direct correlation between 
the latter and the blackbody component {\rc (see first and fourth panels)}. 
\begin{inlinefigure}
\psfig{figure=f3e.eps,height=7.7cm} 
\caption{LECS and MECS spectra of \rx\ for two selected 
phase intervals (0.2--0.4) and (0.8--1.0) as reported in Figure\,2. The  
power--law photon index variation is clearly evident.}
\end{inlinefigure}
  
\vspace{-4mm}
\section{Discussion}
The \BSAX\ spectrum of 1RXS J170849--400910 is well modeled by the sum of 
a (absorbed) relatively steep power--law and a low--energy blackbody. 
Therefore \rx\ is the fourth AXP, after  \uu\ (White et al. 1996; Israel et 
al. 1999b), \ee\ (Corbet et al. 1995; Parmar et al. 1998) and \oo\ 
(Oosterbroek et al. 1998), for which such a spectral decomposition 
has been detected. We note that these sources are also the ones for 
which good spectral data are available. 
This possible decomposition for \rx\ was first suggested by Sugizaki 
(1997) although not statistically significant using the ASCA data. 
The comparison between the \BSAX\ and ASCA observations reveal that 
the source has remained nearly at the same (absorbed) flux level 
($\sim$4.4$\times$10$^{-11}$ erg\,cm$^{-2}$\,s$^{-1}$ and 
$\sim$4.3$\times$10$^{-11}$ erg\,cm$^{-2}$\,s$^{-1}$ in the 0.8--10\,keV 
band for \BSAX\ and ASCA, respectively). {\rc Also the spectral 
parameters (in the  blackbody plus power--law model) are similar to those 
found by ASCA with the exception of the temperature of the blackbody 
which is higher in \BSAX: this is not unusual as \BSAX\ instruments have a 
higher sensitivity below 1\,keV allowing a better evaluation of the 
absorption column and, therefore, of the blackbody component parameters.}

We also tried to fit the spectrum of \rx\ with other models. Among these 
we used two blackbodies.
\begin{inlinefigure}
\psfig{figure=f4e.eps,height=7.9cm} 
\caption{Pulsed fraction of the first two harmonics 
of \src\ as a function of the energy (upper two panels), together with 
the power--law to total absorbed flux ratio and the blackbody to power-low 
unabsorbed flux ratio (lower two panels). See text for details.}
\end{inlinefigure}
 
\noindent Such a decomposition is in agreement 
with accreting, magnetic field decay and cooling models which have been 
invoked to account for the main physical mechanism(s) responsible for 
the X--ray emission of AXPs {\rc (see also Thompson \&  Duncan 1996 and Heyl 
\& Hernquist 1997)}. 
Two (or more) concentric regions with different temperatures are in fact 
expected: the innermost corresponding to the polar caps of the neutron star 
(which is also the hottest) and a larger area around the magnetic 
caps characterised by a lower temperature {\rc (see also DeDeo et al. 2000; 
\"Ozel et al. 2001; Perna et al. 2001)}. Regardless the origin of 
the emission from these 
regions, we note that the size of the blackbodies is always smaller 
than the neutron star surface, even for an unrealistic distance to the AXP 
of $\sim$15\,kpc, with the smallest region 
(R$_{BB}\sim$0.3\,km) being also the hottest ($kT\sim$1.5\,keV). 
The results of a more detailed and systematic spectral and timing study of the 
AXPs observed by \BSAX\ will be reported elsewhere.
 
An analysis of the pulse shape of \rx\ as a function of energy (in the 
0.3--10\,keV) reveals a prominent variation with the minimum at low energies 
corresponding to the maximum at high energies (see also Figure\,1). 
Correspondingly, pulse phase spectroscopy detected 
a significant variation for at least one spectral component, the power--law 
photon index, as a function of phase. The peak in the pulse profile at 
highest energy also corresponds to the lowest value of the photon index, 
similar to what observed in accreting X--ray pulsars (Makishima et al. 1999 
and references therein). Although 
energy--dependent changes in the pulse shape were already 
observed in the past for \uu\ (White et al. 1996; Israel et al. 1999b; 
Paul et al. 2000), the variations of \rx\ are also accompanied 
by a nearly total phase reversal between the low and high energy band, and 
are likely due by a changing power--law slope (as suggested by Sugizaki 
et al. 1997).  
We note that, in the past, the lack of any conspicuous change in the pulse 
profiles as a function of energy and/or pulse phase--resolved spectra of AXPs 
was used to argue against the possibility that these sources are accreting 
X--ray objects.  We finally note that the pulse fraction as 
a function of energy shown in Figure\,4 is not in disagreement with that 
recently reported by \"Ozel et al. (2001) which used a different definition 
of the fractional contribution {\rc (and assuming only one harmonic)} 
to the pulsations.

\rx\ is so far the only AXP for which a sudden spin--up was observed 
(Kaspi et al. 2000). This was interpreted as a glitch similar to that 
observed in the Vela and other young radio pulsars. However, 
we note that glitches could in principle be detected also in accreting 
(spinning--down) X--ray sources with a sufficiently high magnetic field 
strength (in analogy with the known characteristics of radio pulsars) 
if they are in a low noise level phase, as indeed AXPs are 
known to be. 
A way to distinguish, in the near future, whether \rx\ experienced 
a radio pulsar--like glitch or, perhaps, an accreting X--ray 
pulsar--like spin--up behavior would be to accurately monitor the 
period history after the event. 
We note that the lack of the recovery of the \.P value to the pre--event 
one {\rc (in contrast with the known behavior of glitches observed in 
radio pulsars)} would argue against ``magnetar'' models, at least in the 
current formulation, {\rc while its detection might be not conclusive 
for any model (magnetar and accretion).}   

\rx\ is also the first AXP for which spectral changes as a function 
of pulse  phase have been significantly detected.
These spectral/timing properties make \rx\ an especially interesting 
AXP to study.
More sensitive and/or higher throughput observations of \rx\ might yield 
important additional information on the spectral changes causing the pulse 
shape variations and extend the energy range over which the source 
is detected above 10\,keV.

\acknowledgments
This work is supported through CNAA, ASI and Ministero dell'Universit\`a e 
Ricerca  Scientifica e Tecnologica (MURST--COFIN) grants. The authors 
thank the \BSAX\ Mission Planning Team {\rc and an anonymous referee for the 
comments which helped to improve this paper.}

\end{document}